# Informe

# Mejora del proceso de calibración para los WCD del proyecto LAGO usando modelo semieanalítico basado en simulaciones

**Autor: Luis Otiniano Ormachea**

**Investigador: Dirección de Astrofísica de la CONIDA**

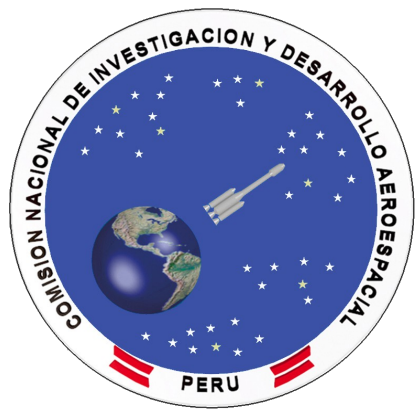

## 1.- Introducción

Con el fin de disminuir costos en la fabricación de los detectores se utilizan tanques de agua comerciales, por lo que no siempre se puede conseguir una óptima separación de las regiones electromagnética y muónica en los detectores. Actualmente para calibrar los detectores de LAGO se suele utilizar el pico secundario del histograma de carga para inferir directamente la distribución de carga del VEM (Muon Vertical Equivalent). El problema también se complica con la perdida de sensibilidad del sensor con el transcurrir del tiempo debido a contaminación del agua del detector.

Sin embargo en [1] medidas directas del VEM utilizando placas de plástico centellador, en detectores prototipo del Observatorio Pierre y un modelo del flujo de muones son usados para reproducir el histograma de carga. En estas medidas el VEM tiene un valor significativamente menor al que se obtendria de ubicar el máximo del histograma de carga.

En esta nota técnica usaremos el proceso inverso, siendo conocido el histograma de carga utilizaremos el modelo de [1] para estimar la distribución de carga del VEM. Luego utilizaremos esta distribución del VEM y una simulación de flujo de secundarios de una hora para simular la componente muónica total del histograma.

## 2.- Estudio analítico

Se desarrolla en esta parte un re-escalamiento del histograma de carga dejado por las partícuals al atravezar el detector, usualmente usado en la calibración del mismo para estimar el valor del VEM. El reescalamiento permite una mejor estimación del VEM como demostraremos

De [1] tenemos que dado un detector de radio **r** *y altura **h*** para un detector en que se cumpla que **2r>>h**

1. $dn(\theta) = 2\pi^2 2r^2 t \Phi_0 \cos^3(\theta) sen(\theta) d\theta$

donde t es el tiempo y se ha usado que la distribución del flujo de muones es de la forma:

2. $\Phi = \Phi_0 \cos^2(\theta)$

y en esta aproximación el camino recorrido por un muon que atraviesa el detector *l* sería :

3. $l = h/\cos(\theta)$

luego

4. $\dfrac{dn(l)}{t\Phi_0} = 2\pi^2 r^2 h^4 \dfrac{dl}{l^5}$

De otro lado si asumimos que la distribución de carga del VEM es de tipo gaussiana normalizada:

5. $V(q) = N_{(s0)} e^{\dfrac{(q-q0)^2}{2s0^2}}$

Asumiremos como en 1 que s0 y q0 dependen linealmente de L entonces se puede escribir

6. $V(q,l) = N_{(s0 \times l/h)} e^{\frac{(q - q0 \times l/h)^2}{2(s0 \times l/h)^2}} = N_{(s0)} h/l e^{\frac{(q \times h/l - q0)^2}{2(s0)^2}} = h/l V(qh/l)$

La carga total que ingresa al detector por estos muones esta dada por:

7. $F(q) = \int_0^{\sqrt{4r^2 + h^2}} V(q,l) \frac{dn(l)}{dl} dl$

utilizando 6 y 4 y con el cambio de variable $l \rightarrow qh/x$ tenemos y restringiéndonos a muones que recorran caminos en el detector mayores a los verticales $l \in [h, \sqrt{4r^2 + h^2}]$, lo que es valido en la aproximación de r>>h

8. $F(q) = (2\pi^2 2r^2 t \Phi_0) N_{(s0)} (-q^5 h) \int_q^{q\frac{h}{\sqrt{4r^2+h^2}}} e^{\frac{(x-q0)^2}{2s0^2}} x^4 dx$

Reacomodando términos y derivando con respecto a q tenemos:

9. $5q^4 F(q) + q^5 F'(q) = A \left( e^{\frac{(x-q0)^2}{2s0^2}} x^4 \right)_{qh}^{q\frac{h}{\sqrt{4r^2+h^2}}}$

Evaluando la integral finalmente tenemos:

10. $F(q) + \frac{q}{5} F'(q) = \frac{A}{5} \left\{ e^{\frac{(q-q0)^2}{2s0^2}} - \left( \frac{h}{\sqrt{4r^2+h^2}} \right)^4 e^{\frac{(q\frac{h}{\sqrt{4r^2+h^2}} - q0)^2}{2s0^2}} \right\}$

Notamos que si nuestro detector cumple **2r>>h** el segundo miembro del lado derecho se puede despreciar y que podemos estimar la distribución del VEM directamente del histograma de carga con una corrección. Sin embargo necesitamos una buena separación de las componentes muónicas y electromagnéticas para lograr esto.

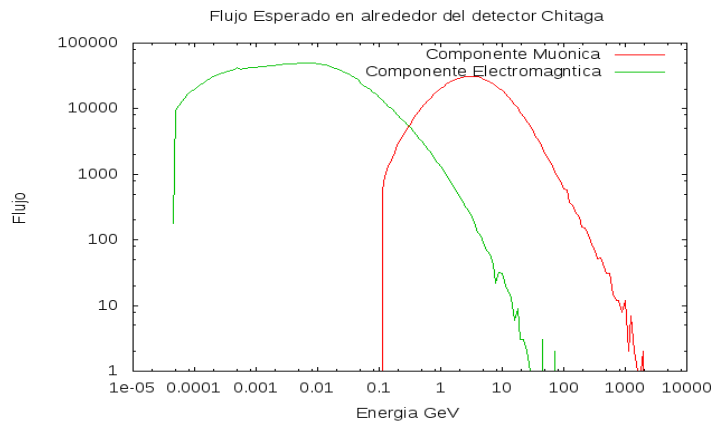

*Figura 1: Distribución de Energía del flujo de partículas esperados en el entorno del detector Chitaga simulados con CORSIKA.*

## 3.- Distribución de Muones en Chitaga

Utilizando una simulación de CORSIKA [3] y el set de herramientas del proyecto lago LAGO-CORSIKA [4] que permite estudiar el entorno de partículas secundarias que esperaríamos sobre un metro cuadrado alrededor del detector, encontramos que el rengo de energías de las componentes muónicas estarían por los GeV. (Ver Figura 1). Como en promedio los muones pierden 2 MeV/cm al atravesar la materia podemos asumir que todos los muones atraviezan directamente al detector, dejando de lado el proceso de decaimiento.

## 4.- Histograma de Carga de Chitaga

Como ejemplo tomaremos a Chitaga un detector de la Universidad Industrial de Santander a cuyos datos se tiene acceso via el repositorio de LAGO. En la figura 2 mostramos el histograma de carga de 1 hora del detector. Las componentes muónica y electromagnética esta superpuestas, aplicamos la selección descrita en [2], que utiliza el hecho de que la señal muónica presenta picos más pronunciados que la componente electromagnética. (Esencialmente dados dos pulsos consecutivos se descarta el primero si su pico es menor). En la figura 2. también graficamos el lado izquierdo de la ecuación 10, calculado a partir del histograma de máxima popularidad muónica, luego de un suavizar la derivada.

Aunque para Chitaga r=100 cm, y h=90cm y no cumplan con la condición, la respuesta al flujo cerca de la vertical se esperaría ser parecida a la de un detector muy grande, al menos para la posición el pico muónico, lo primero que notamos es que el pico tiene un valor menor al pico en el histograma de la selección muónica.

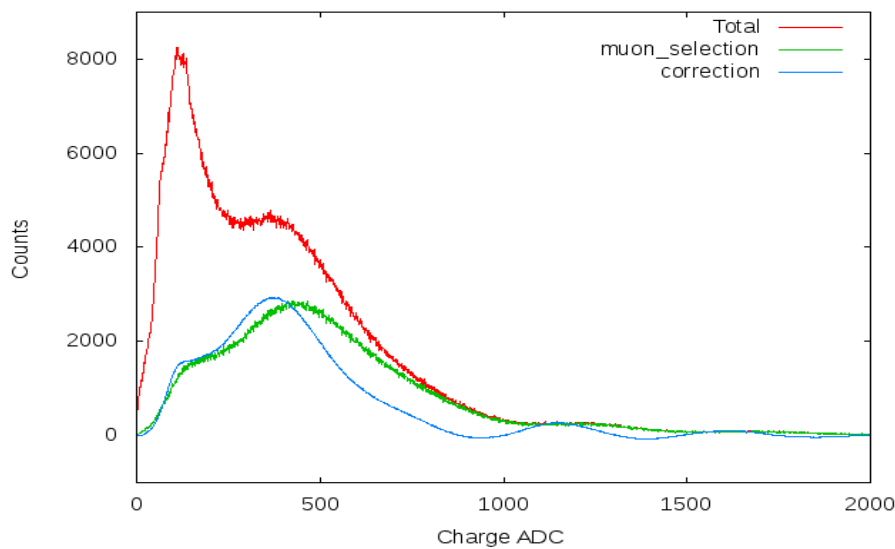

*Figura 2: Histograma carga de 1 hora de Chitaga (rojo), selección muónica (verde) y corrección a la selección muónica utilizando la ecuación 10 (azul).*

Haciendo un ajuste de parámetros en la ecuación 10 en una rango alrededor del máximo obtenemos *$s0=383$ , $q0=383.45$ ADC*. Con el fin de validar el método utilizaremos la ecuación 6 para estimar la contribución muónica total de la simulación de 1 hora de secundarios sobre Chitaga. Para esto asumimos que los muones atravezarian el detector en línea recta, en la figura 3 mostramos el histograma de distancias atravezadas por los muones secundarios simulados distribuyéndolos al azar sobre un cuadrado de lado iguala 6 veces el radio del detector, esto fue realizado con un pequeño código en c++ el cual simplemente mide las distancias que recorrerian los muones simumldos con Corsika si aparecieran en una región cuadrada sobre el detector. En la figura 4 mostramos los resultados.

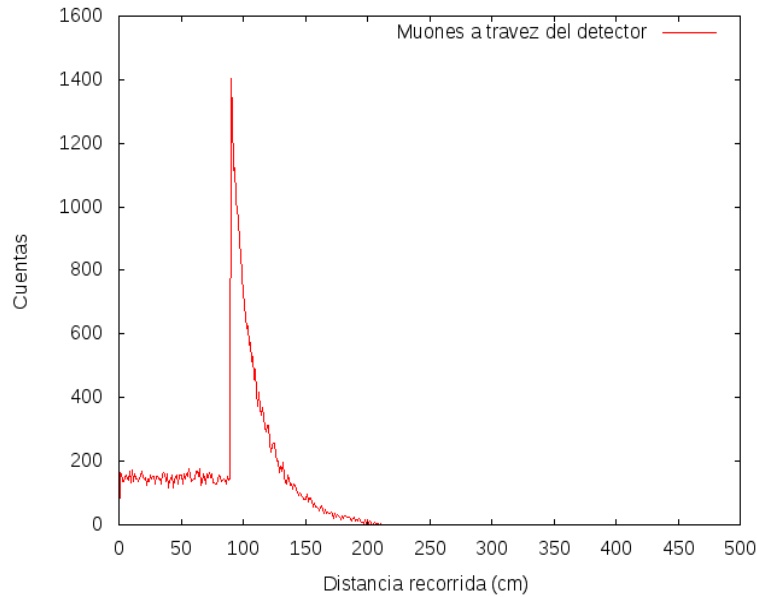

*Figura 3: Histograma del camino recorrido por muones simulados que atravezarían el detector.*

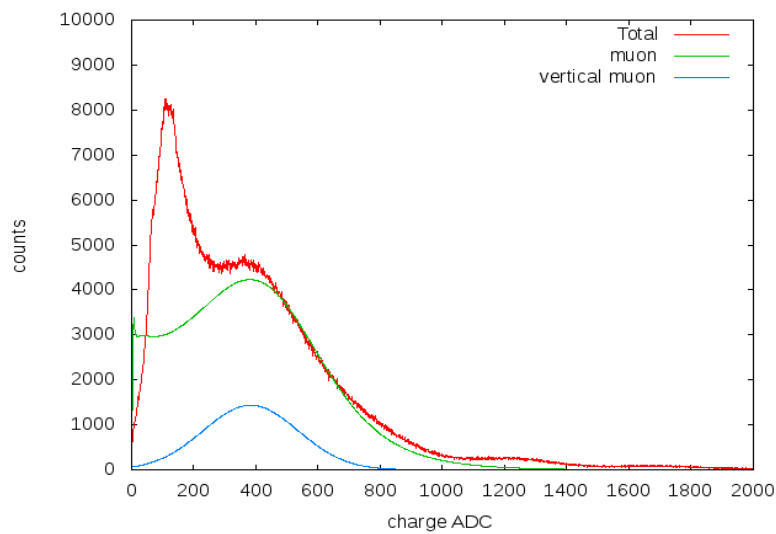

*Figura 4: Histograma de carga del detector (rojo) comparado con respuesta simulada del detector al flujo de muones (verde) utilizando la ecuación 6 el camino recorrido por muones simulados que atravesarían el detector. También se muestra la respuesta simulada a muones verticales (azul)*

## 5.- Discusión

Si bien se ha podido corregir la posición del máximo es necesario realizar esta estimación con otros detectores y validarla con simulaciones o con medidas de muones verticales. En donde se cuenta con detectores del proyecto Escaramujo eso es posible y actualmente estamos trabajando para crear un circuito de acople entre el los detectores de LAGO y los de Escaramujo con la idea de validar este método.

También es necesario realizar un trabajo similar para estimar la componente electromagnética, para ello necesitamos un subconjunto del histograma donde solo se cuente con señal electromagnética, el espectro de electrones provenientes del decaimiento muónico nos permitiría esto. Pero esta parte escapa a los fines de este informe al intentar una calibración puramente electromagnetica.

Finalmente podemos decir que el $q_0=383.45$ ADC corresponde a una deposicion de 180 MeV recordando que un muon deposita aproximandamente 2 MeV por cm y el detector tiene una altura de 180 cms.

## 6.- Bibliografía